\documentclass[fleqn,usenatbib]{mnras}

\usepackage[T1]{fontenc}

\DeclareRobustCommand{\VAN}[3]{#2}
\let\VANthebibliography\thebibliography
\def\thebibliography{\DeclareRobustCommand{\VAN}[3]{##3}\VANthebibliography}

\usepackage{graphicx}	
\usepackage{amsmath}	
\usepackage{amssymb}	

\title
[183 GHz H$_{2}$O maser emission in Superantennae]
{ALMA detection of millimetre 183 GHz H$_{2}$O maser emission 
in the Superantennae galaxy at z $\sim$ 0.06}

\author[M. Imanishi et al.]
{Masatoshi Imanishi, $^{1,2}$\thanks{E-mail: masa.imanishi@nao.ac.jp}
Yoshiaki Hagiwara, $^{3}$
Shinji Horiuchi, $^{4}$
Takuma Izumi$^{1,2}$, 
and 
Kouichiro Nakanishi $^{1,2}$
\\
$^{1}$National Astronomical Observatory of Japan, National Institutes 
of Natural Sciences (NINS), 2-21-1 Osawa, Mitaka, Tokyo 181-8588, Japan \\
$^{2}$Department of Astronomy, School of Science, The Graduate University 
for Advanced Studies, SOKENDAI, Mitaka, Tokyo 181-8588, Japan \\
$^{3}$Natural Science Laboratory, Toyo University, 5-28-20 Hakusan, 
Bunkyo-ku, Tokyo 112-8606, Japan \\
$^{4}$CSIRO Astronomy and Space Science, Canberra Deep Space 
Communications Complex, PO Box 1035, Tuggeranong, ACT 2901, Australia \\
}

\date{Accepted XXX. Received YYY; in original form ZZZ}

\pubyear{2020}

\begin{document}
\label{firstpage}
\pagerange{\pageref{firstpage}--\pageref{lastpage}}
\maketitle

\begin{abstract}
We present the results of ALMA band-5 ($\sim$170 GHz) 
observations of the merging ultraluminous infrared galaxy, 
the ``Superantennae'' (IRAS 19254$-$7245) at $z=$0.0617, which has been 
diagnosed as containing a luminous obscured active galactic nucleus (AGN).
In addition to dense molecular line emission (HCN, HCO$^{+}$, and 
HNC $J$ = 2--1), 
we detect a highly luminous ($\sim$6$\times$10$^{4}L_{\odot}$) 183 GHz 
H$_{2}$O 3$_{1,3}$--2$_{2,0}$ emission line.
We interpret the strong H$_{2}$O emission as largely originating in maser 
amplification in AGN-illuminated dense and warm molecular gas, based on 
(1) the spatially compact ($\lesssim$220 pc) nature of 
the H$_{2}$O emission, unlike spatially resolved ($\gtrsim$500 pc) 
dense molecular emission, and  
(2) a strikingly different velocity profile from, and (3) 
significantly elevated flux ratio relative to, dense 
molecular emission lines.
H$_{2}$O maser emission, other than the widely studied 22 GHz 
6$_{1,6}$--5$_{2,3}$ line, has been expected to provide important information on 
the physical properties of gas in the vicinity of a central mass-accreting 
supermassive black hole (SMBH), because of different excitation energy.
We here demonstrate that with highly sensitive ALMA, millimetre 
183 GHz H$_{2}$O maser detection is feasible out to $>$270 Mpc, 
opening a new window to scrutinize molecular gas properties 
around a mass-accreting SMBH far beyond the immediately local universe.
\end{abstract}




\section{Introduction} 

Water (H$_{2}$O) is an abundant molecule in the universe and has been 
detected in many active galaxies \citep{yan13}.
The rotational energy levels of H$_{2}$O are more complex than those of simple 
molecules (e.g., CO, HCN) and many rotational transition lines are 
found in the far-infrared (70--300 $\mu$m), (sub)millimetre (0.3--10 mm), 
and centimetre ($>$1 cm) wavelength ranges.
In dense and warm molecular gas, population inversion can occur 
for a number of H$_{2}$O rotational transitions through collisional 
excitation and/or infrared radiative pumping \citep[e.g.,][]{yat97,gon10}.
This population inversion can amplify background radiation and 
these H$_{2}$O emission lines can be extremely bright through maser 
phenomena.
The luminous megamaser ($>$10$L_{\odot}$) emission line 
of ortho-H$_{2}$O 6$_{1,6}$--5$_{2,3}$ at rest frequency $\nu_{\rm rest}$ 
$\sim$ 22 GHz (1.35 cm) has been detected in galaxy nuclei, 
mostly obscured active galactic nuclei (AGNs) 
\citep[e.g.,][]{bra96,gre03b,bra04,hen05,kon06a,kon06b}, 
out to z $\sim$ 0.66 for an unlensed AGN \citep{bar05} and z $\sim$ 2.6 
for a lensed AGN \citep{imp08}.
Dense and warm molecular gas in the vicinity of a luminous AGN is 
a plausible site for this H$_{2}$O megamaser emission \citep{neu94,mal02}, 
which often shows brighter blueshifted and redshifted components than the 
systemic velocity component caused by a highly edge-on rotating disc. 
Because high-spatial-resolution very-long-baseline interferometry (VLBI) 
observations are possible 
at centimetre wavelengths, detailed spatially resolved dynamical studies of 
the bright 22 GHz H$_{2}$O megamaser emission can be an excellent probe of 
the surrounding mass distribution.
Such VLBI observations provided the first convincing evidence of 
the presence of a supermassive black hole (SMBH) and its precise mass 
measurement in the nearby AGN NGC 4258 at $\sim$7 Mpc \citep{miy95}.
Subsequent centimetre VLBI observations also revealed the dynamical 
properties of 
H$_{2}$O megamaser-emitting gas in close proximity to central 
mass-accreting SMBHs and constrained SMBH masses in other nearby 
well-studied AGNs, including 
NGC 4945 at $\sim$4 Mpc \citep{gre97b}, 
Circinus at $\sim$4 Mpc \citep{gre03a}, and others 
\citep[e.g.,][]{kuo11}.

Theoretically, other H$_{2}$O rotational transition lines in the 
(sub)millimetre wavelength range can also be extremely bright as a result 
of the maser 
phenomena caused by population inversion in dense and warm molecular gas.
Examples include the 3$_{1,3}$--2$_{2,0}$ transition line of para-H$_{2}$O 
at $\nu_{\rm rest}$ $\sim$ 183 GHz,  
10$_{2,9}$--9$_{3,6}$ line of ortho-H$_{2}$O at $\nu_{\rm rest}$ $\sim$ 321 GHz, 
and 5$_{1,5}$--4$_{2,2}$ line of para-H$_{2}$O at $\nu_{\rm rest}$ $\sim$ 325 GHz 
\citep[e.g.,][]{deg77,neu91,yat97}.
Observing multiple H$_{2}$O megamaser emission lines will enable us to 
constrain the physical properties of the innermost AGN-illuminated gas 
around a mass-accreting SMBH \citep{mal02,hag13}, 
because excitation energy levels differ distinctly between different lines.

With the advent of highly sensitive (sub)millimetre observing facilities, 
including the Atacama Large Millimeter/submillimeter Array (ALMA), 
detection of (sub)millimetre H$_{2}$O emission 
(183 GHz H$_{2}$O 3$_{1,3}$--2$_{2,0}$ and/or 
321 GHz H$_{2}$O 10$_{2,9}$--9$_{3,6}$ lines) has been reported 
in the very nearby obscured AGNs Circinus ($\sim$4 Mpc), 
NGC 4945 ($\sim$4 Mpc), and NGC 3079 ($\sim$16 Mpc)
\citep{hum05,hag13,hag16,pes16,hum16}.
These detected (sub)millimetre H$_{2}$O emission lines can be explained by 
maser phenomena based mainly on high luminosity and velocity profiles 
similar to the 22 GHz H$_{2}$O megamaser emission previously 
detected by VLBI observations.
The 183 GHz H$_{2}$O emission line was also detected in the infrared 
luminous merging galaxy Arp 220 at z $\sim$ 0.018 ($\sim$80 Mpc) 
\citep{cer06}.
While \citet{cer06} interpreted the emission as masers originating from nuclear 
star-forming regions, \citet{hag16} and \citet{gal16} 
argued that it is more likely to be thermal emission.
\citet{kon17} later detected the 183 GHz and 325 GHz H$_{2}$O emission 
lines in Arp 220 and preferred the hypothesis of masers from star-forming 
regions (not from an AGN).
Despite the potential importance of (sub)millimetre H$_{2}$O megamaser 
emission from AGN-illuminated molecular gas, it remains largely 
unexplored beyond the immediately local universe at $>$20 Mpc. 

In this paper, we report the detection of notably luminous 
183 GHz H$_{2}$O emission that is interpreted as being of maser origin, 
in the obscured-AGN-hosting infrared luminous merging galaxy,  
the ``Superantennae'' at $z =$ 0.0617.
Adopting H$_{0}$ $=$ 71 km s$^{-1}$ Mpc$^{-1}$, $\Omega_{\rm M}$ = 0.27, 
and $\Omega_{\rm \Lambda}$ = 0.73, its luminosity distance is 273 Mpc 
and 1 arcsec corresponds to 1.2 kpc.

\section{Target}

The ``Superantennae'' (IRAS 19254$-$7245) is a merging galaxy in the 
southern hemisphere (declination $\sim$ $-$72$^{\circ}$), consisting of 
two galaxy nuclei with a separation of $\sim$8'' ($\sim$10 kpc) along the 
north--south direction and prominent long ($>$300 kpc) tidal tails 
\citep{mel90,mir91,duc97}. 
It has infrared luminosity $L_{\rm IR}$ $\sim$ 1 $\times$ 10$^{12}L_{\odot}$ 
\citep{mel90}, belonging to the class of ultraluminous infrared galaxies 
(ULIRGs) \citep{san96}.
The Superantennae is a scaled-up version of the classical merging galaxy  
``Antennae'' (NGC 4038/9), with $>$3 times longer tidal tails and 
$\sim$10 times higher infrared luminosity \citep{san96}.
The southern nucleus is classified as an obscured AGN through optical 
spectroscopy \citep{mir91,col91,duc97,kew01}, and signatures of a luminous 
obscured AGN behind a large column density of material were also  
found based on $>$2 $\mu$m infrared spectroscopy 
\citep{gen98,van02,ris03,reu07,ima08,nar09,nar10} and 
$>$2 keV X-ray spectroscopy \citep{ima99,pap00,bra03,bra09,jia12}.

\section{Observations and Data Analysis} 

The 183 GHz H$_{2}$O line data of the Superantennae were 
obtained in our ALMA Cycle 5 program 2017.1.00022.S (PI = M. Imanishi) 
as part of band-5 (163--211 GHz) observations of $z <$ 0.15 ULIRGs 
with dense molecular tracers 
HCN $J$ = 2--1 ($\nu_{\rm rest}$ = 177.261 GHz), 
HCO$^{+}$ $J$ = 2--1 ($\nu_{\rm rest}$ = 178.375 GHz), and 
HNC $J$ =2--1 ($\nu_{\rm rest}$ = 181.325 GHz) 
(M. Imanishi et al. 2021 in preparation).
The H$_{2}$O 3$_{1,3}$--2$_{2,0}$ line ($\nu_{\rm rest}$ = 183.310 GHz) data 
were simultaneously obtained with the HNC $J$ = 2--1 line data on 
2018 September 18 (UT) with 43 antennas with 15--1398 m baselines.
The bandpass, flux, and phase calibrator were J1617$-$5848, J1617$-$5848, 
and J1837$-$7108, respectively. 
The net on-source integration time was 27 min.
HCN $J$ = 2--1 and HCO$^{+}$ $J$ = 2--1 line data were taken separately 
on 2018 September 18 (UT) with the same antenna numbers, baselines, and 
calibrators. 
The net on-source integration time was 6 min.

We used CASA (https://casa.nrao.edu) for the reduction of data calibrated 
and provided by ALMA. 
We selected channels that displayed no obvious emission lines to determine 
the continuum level, and then subtracted it using the CASA task ``uvcontsub''.
The ``clean'' task was applied with Briggs weighting 
(robust $=$ 0.5 and gain $=$ 0.1) for both the continuum-only and 
continuum-subtracted molecular line data, with a pixel scale of 
0$\farcs$05 pixel$^{-1}$.
The synthesized beam size was $\sim$0$\farcs$55 $\times$ 0$\farcs$35 
($\sim$650 pc $\times$ 400 pc).

\section{Results} 

Continuum emission is clearly detected at International Celestial 
Reference System (ICRS) coordinates of 
(19$^{\rm h}$ 31$^{\rm m}$ 21.43$^{\rm s}$, $-$72$^{\circ}$ 39$'$  21.5$''$) 
(i.e., southern nucleus) and is displayed as contours in Figure 1 (Top).
In Figure 1 (Top), integrated intensity (moment 0) maps of the 183 GHz 
H$_{2}$O, HNC $J$ = 2--1, HCO$^{+}$ $J$ = 2--1, and HCN $J$ = 2--1 lines 
are shown as images. 
These molecular emission line peak positions spatially agree with the 
continuum peak.
Table 1 (columns 1--4) summarizes these molecular emission line properties.

\begin{figure*}
\begin{center}
\includegraphics[angle=0,scale=.195]{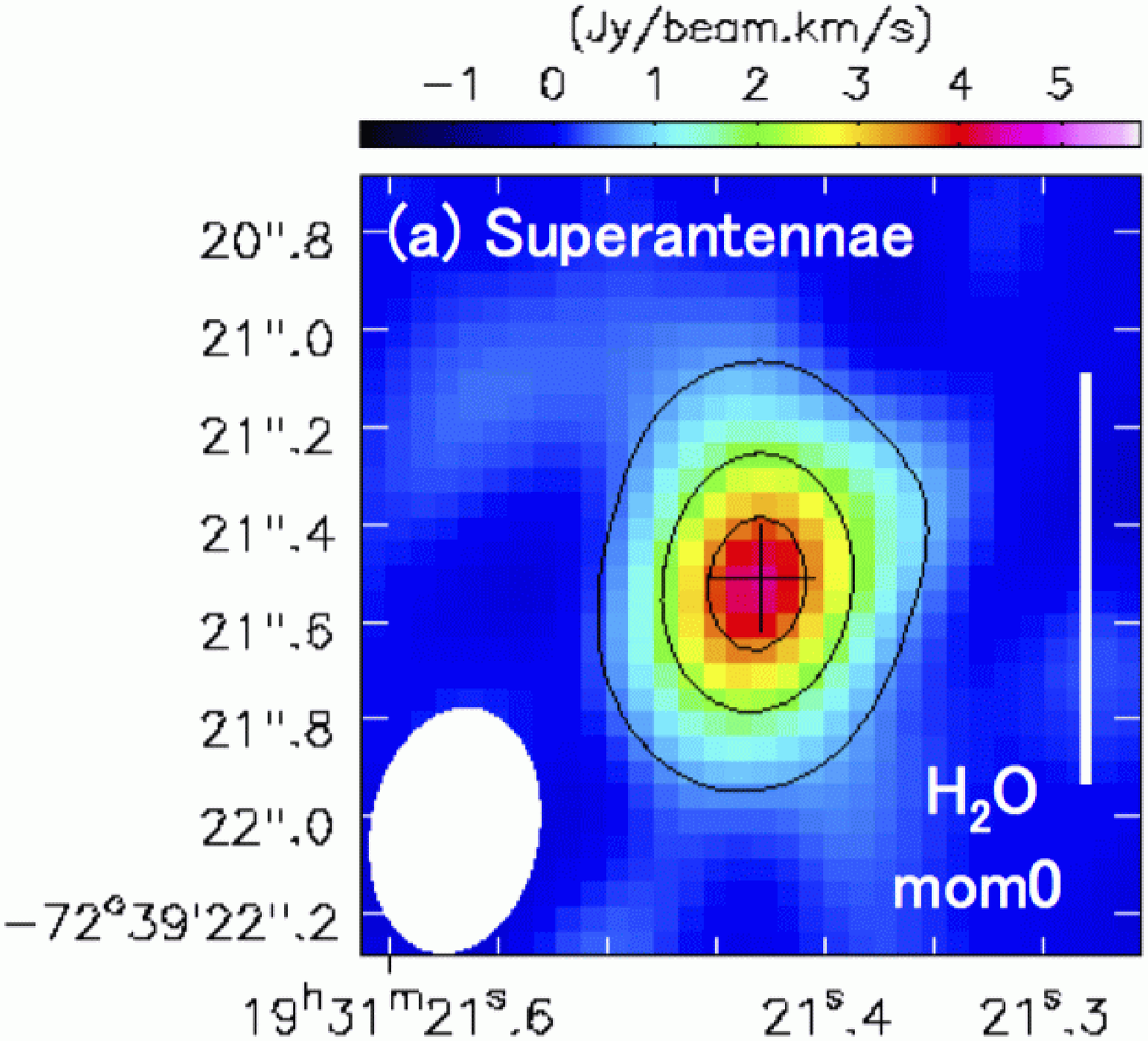} 
\includegraphics[angle=0,scale=.195]{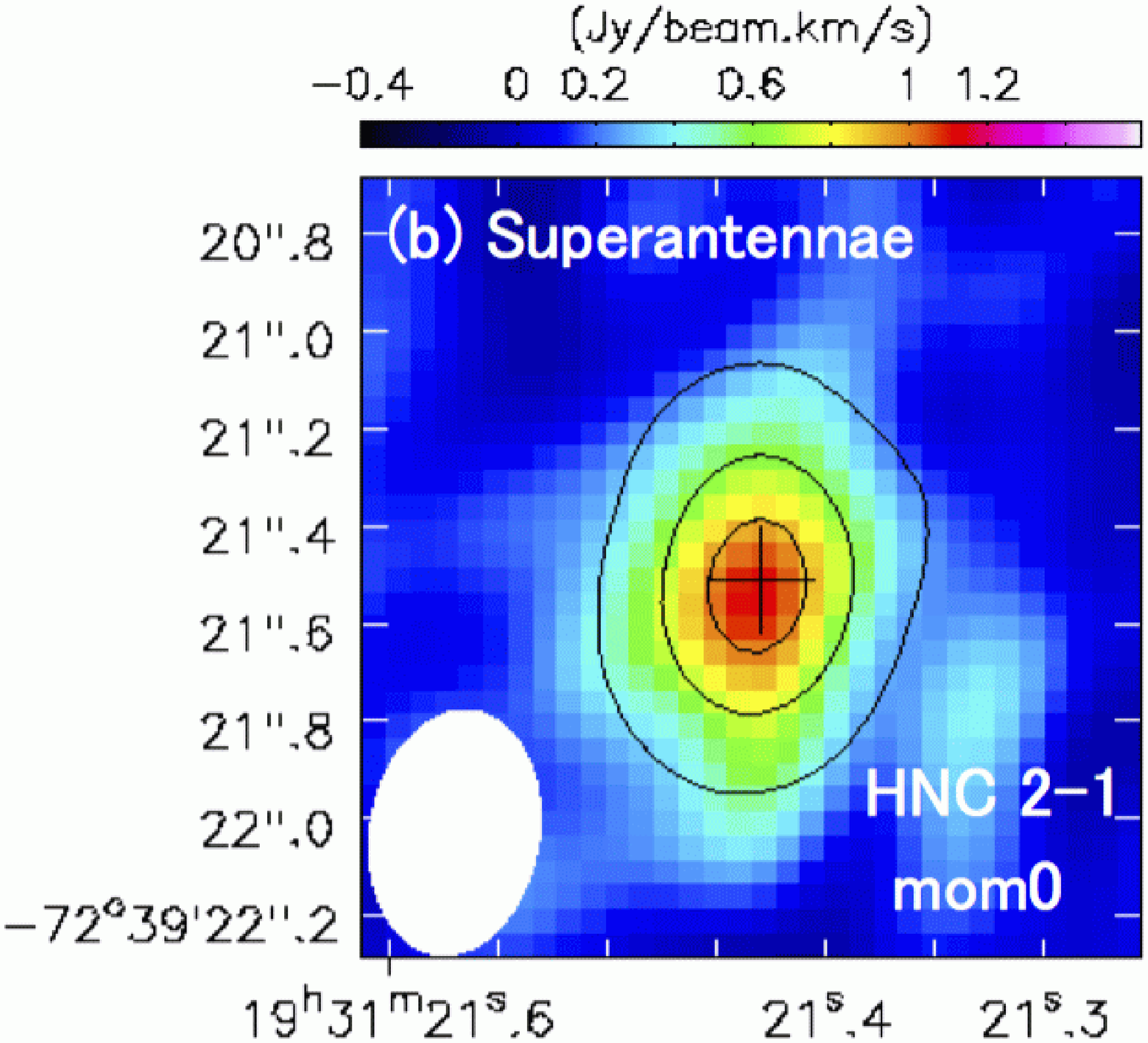} 
\includegraphics[angle=0,scale=.195]{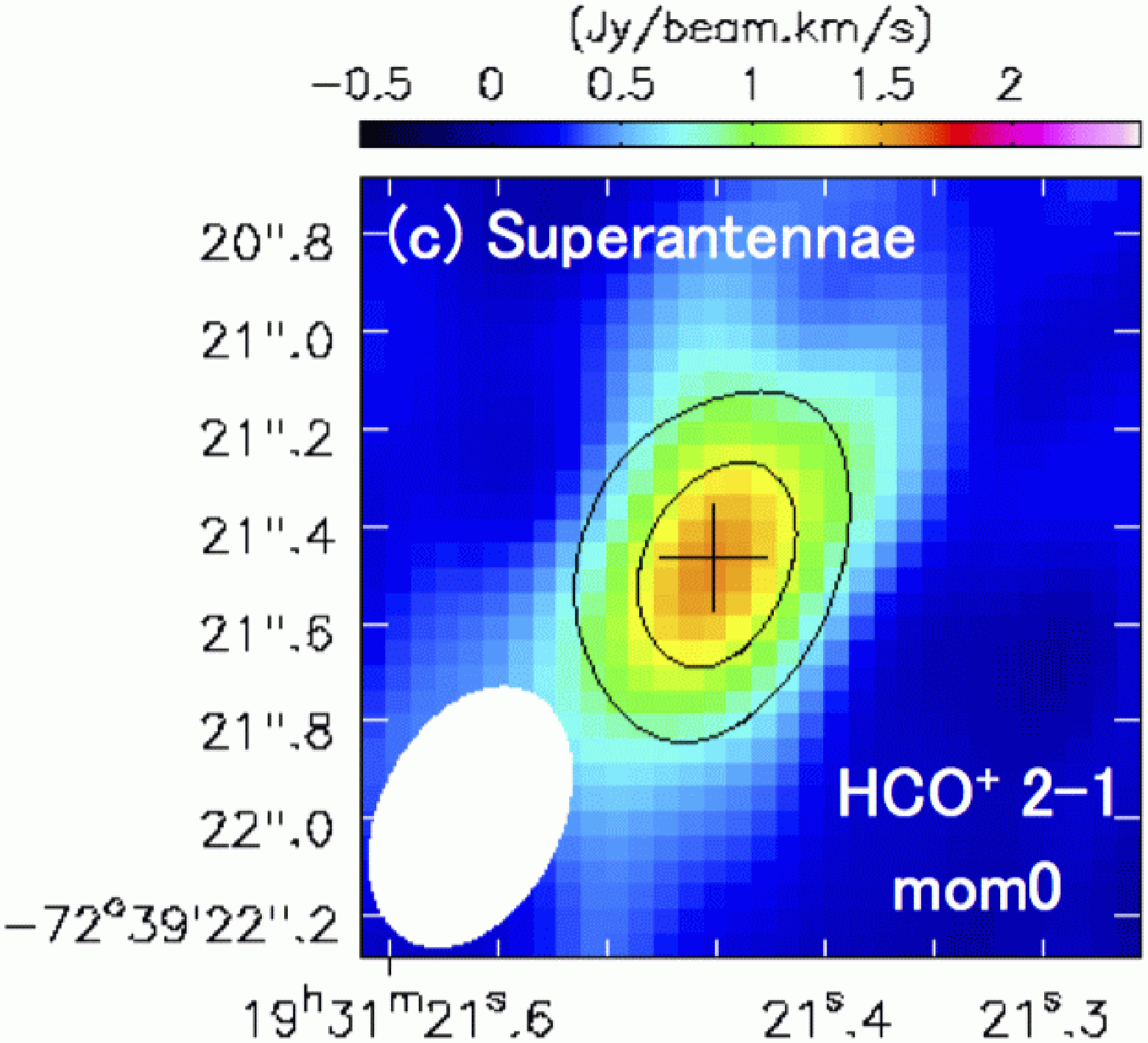} 
\includegraphics[angle=0,scale=.195]{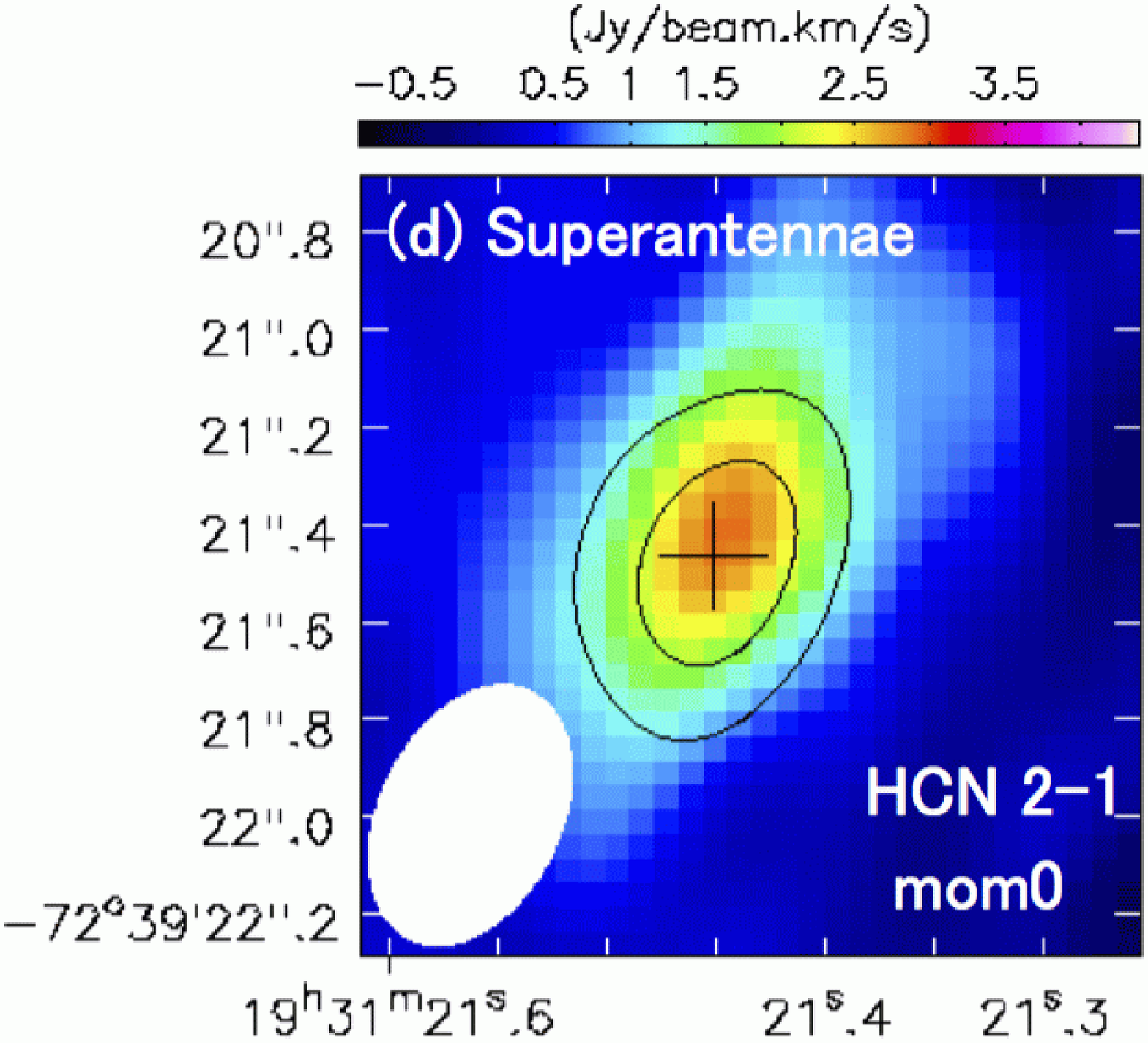} \\
\vspace{0.2cm}
\includegraphics[angle=0,scale=.195]{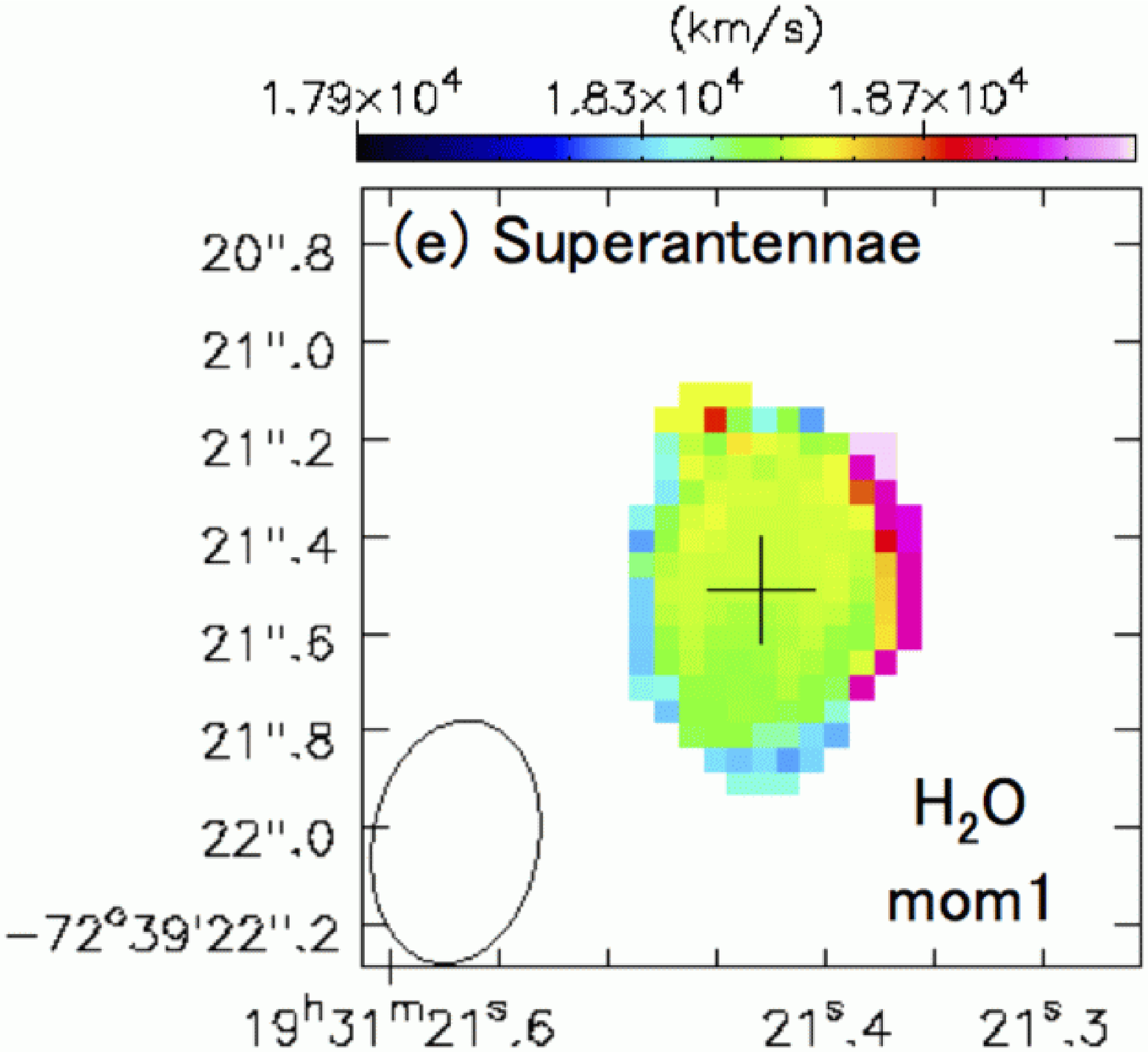} 
\includegraphics[angle=0,scale=.195]{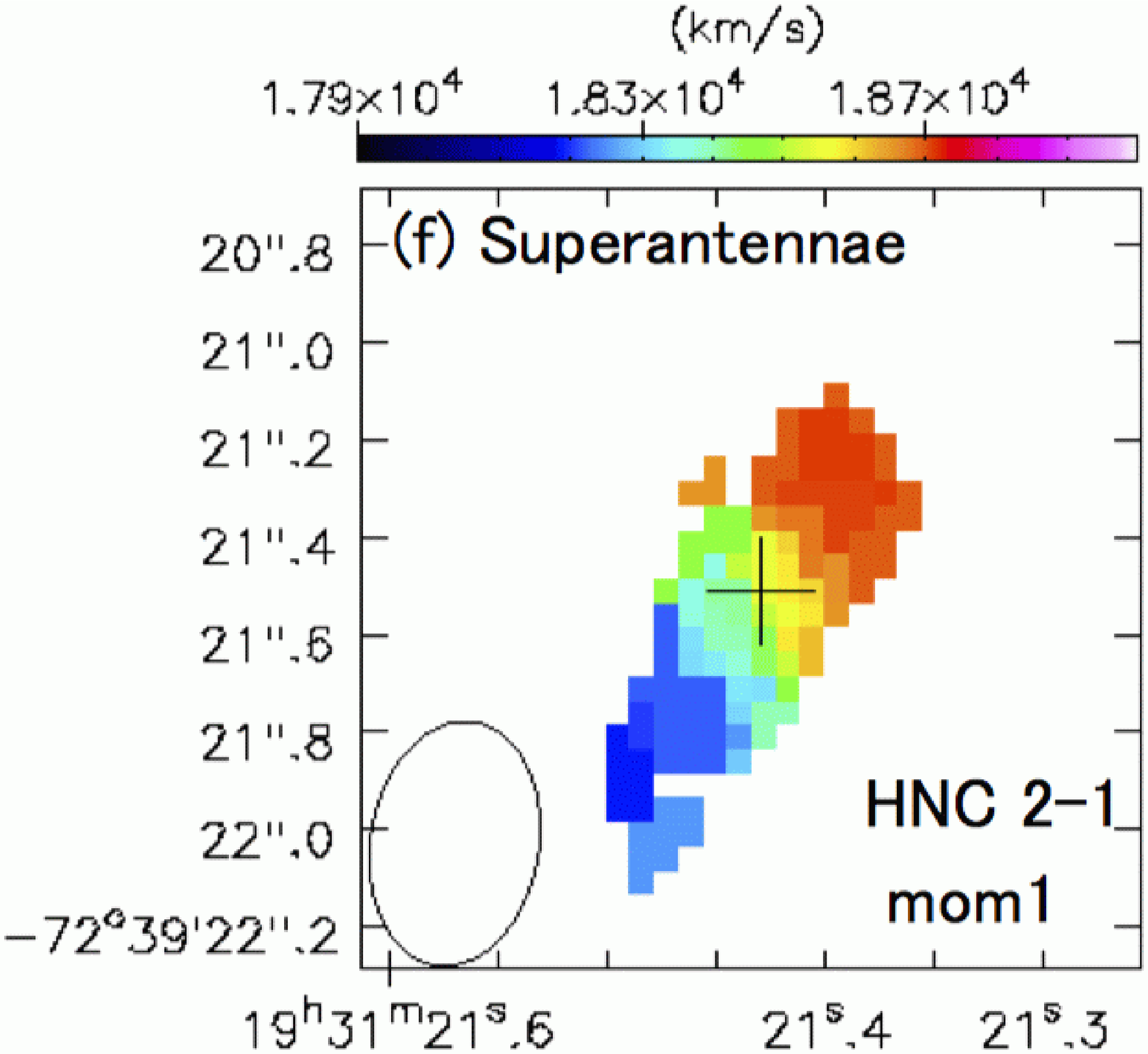} 
\includegraphics[angle=0,scale=.195]{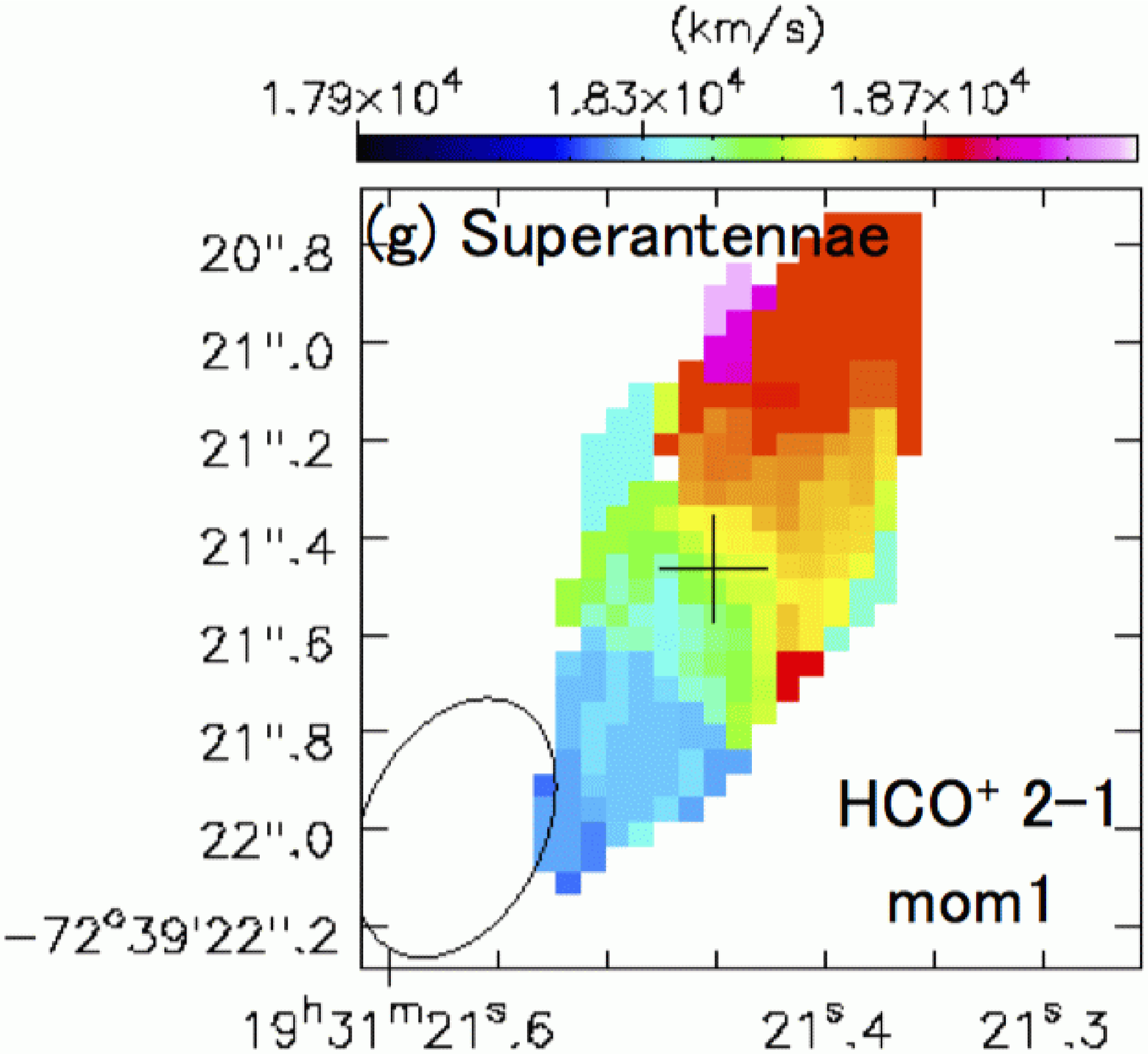} 
\includegraphics[angle=0,scale=.195]{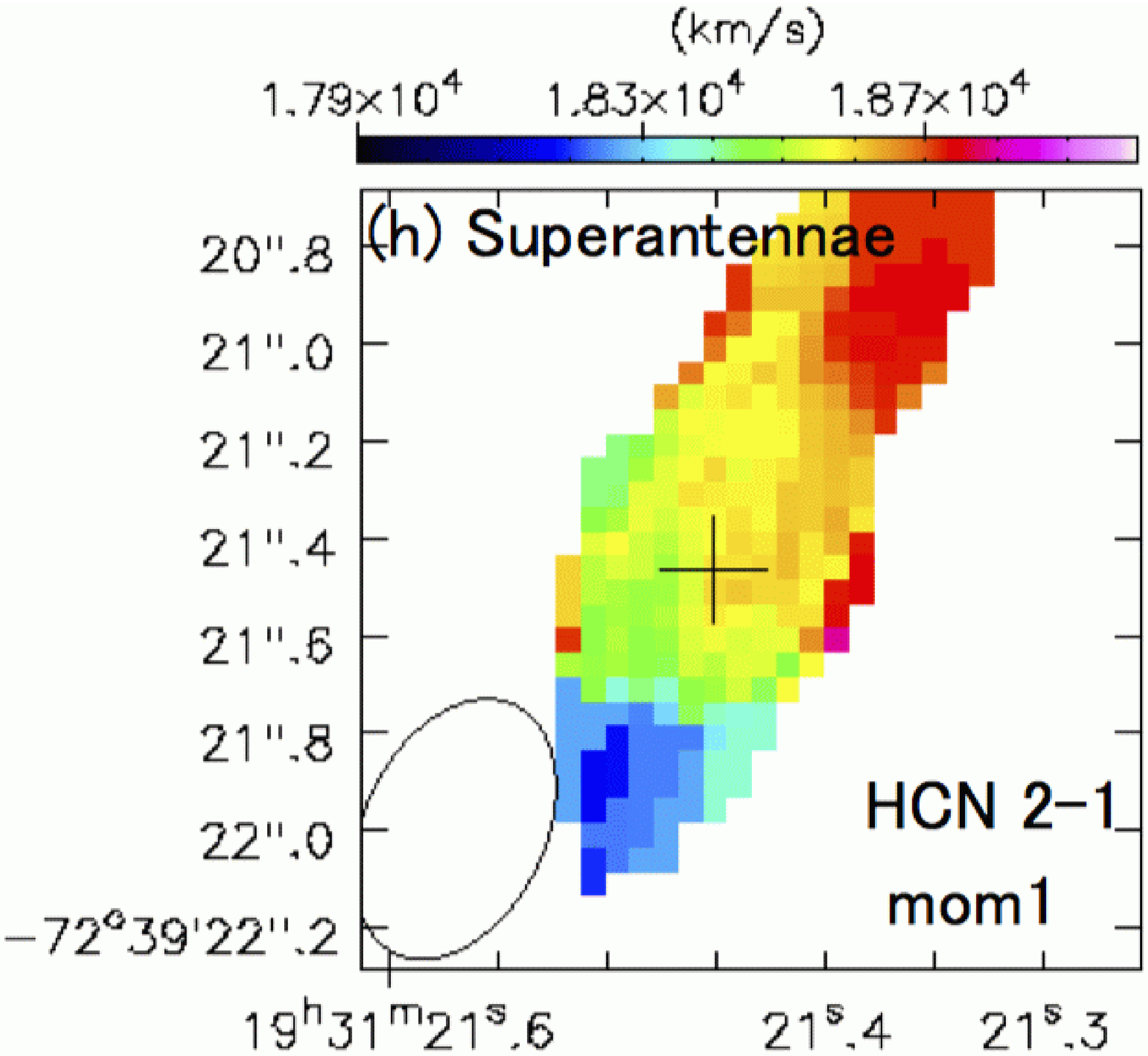} \\
\vspace{0.2cm}
\includegraphics[angle=0,scale=.195]{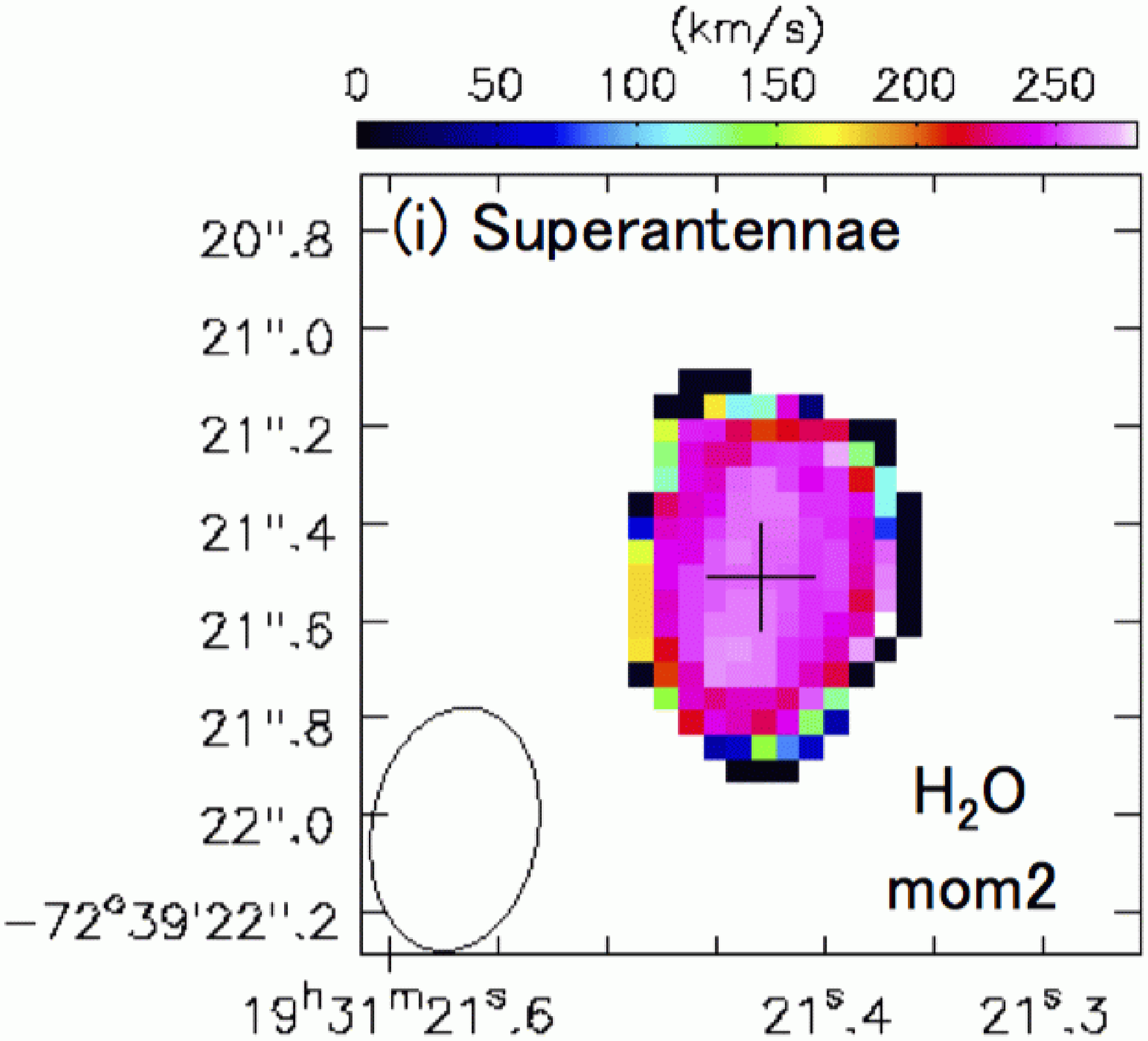} 
\includegraphics[angle=0,scale=.195]{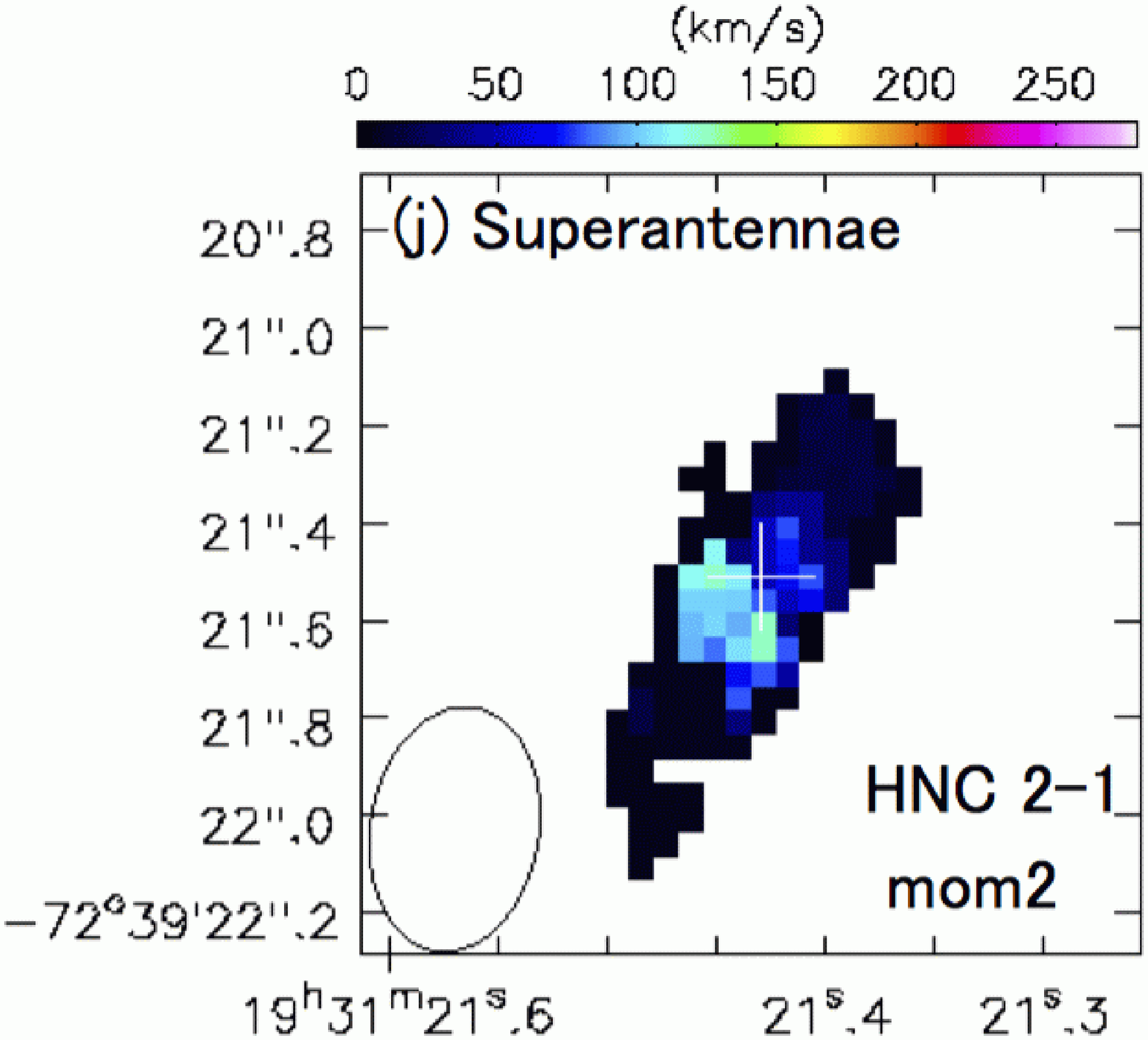} 
\includegraphics[angle=0,scale=.195]{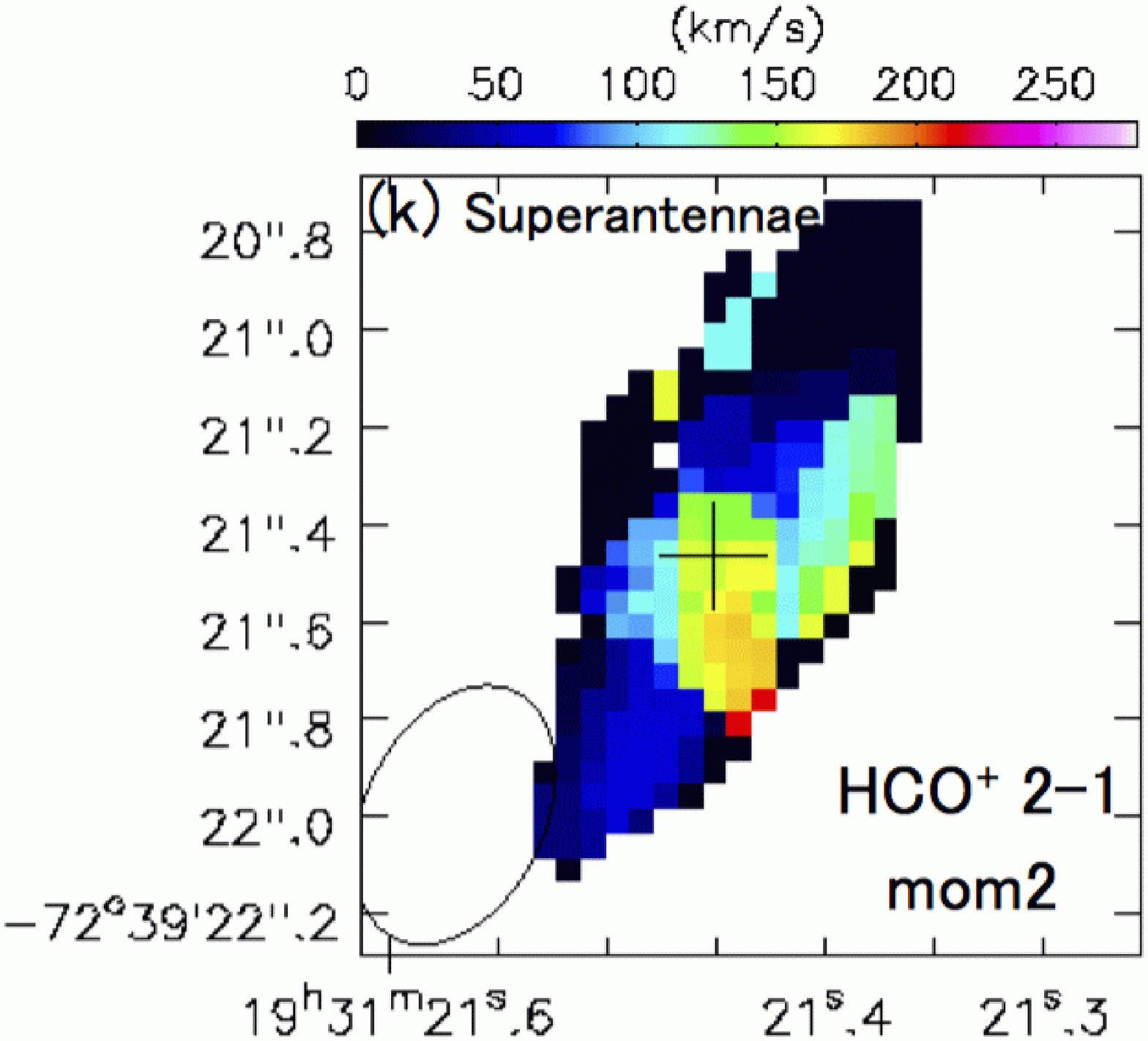} 
\includegraphics[angle=0,scale=.195]{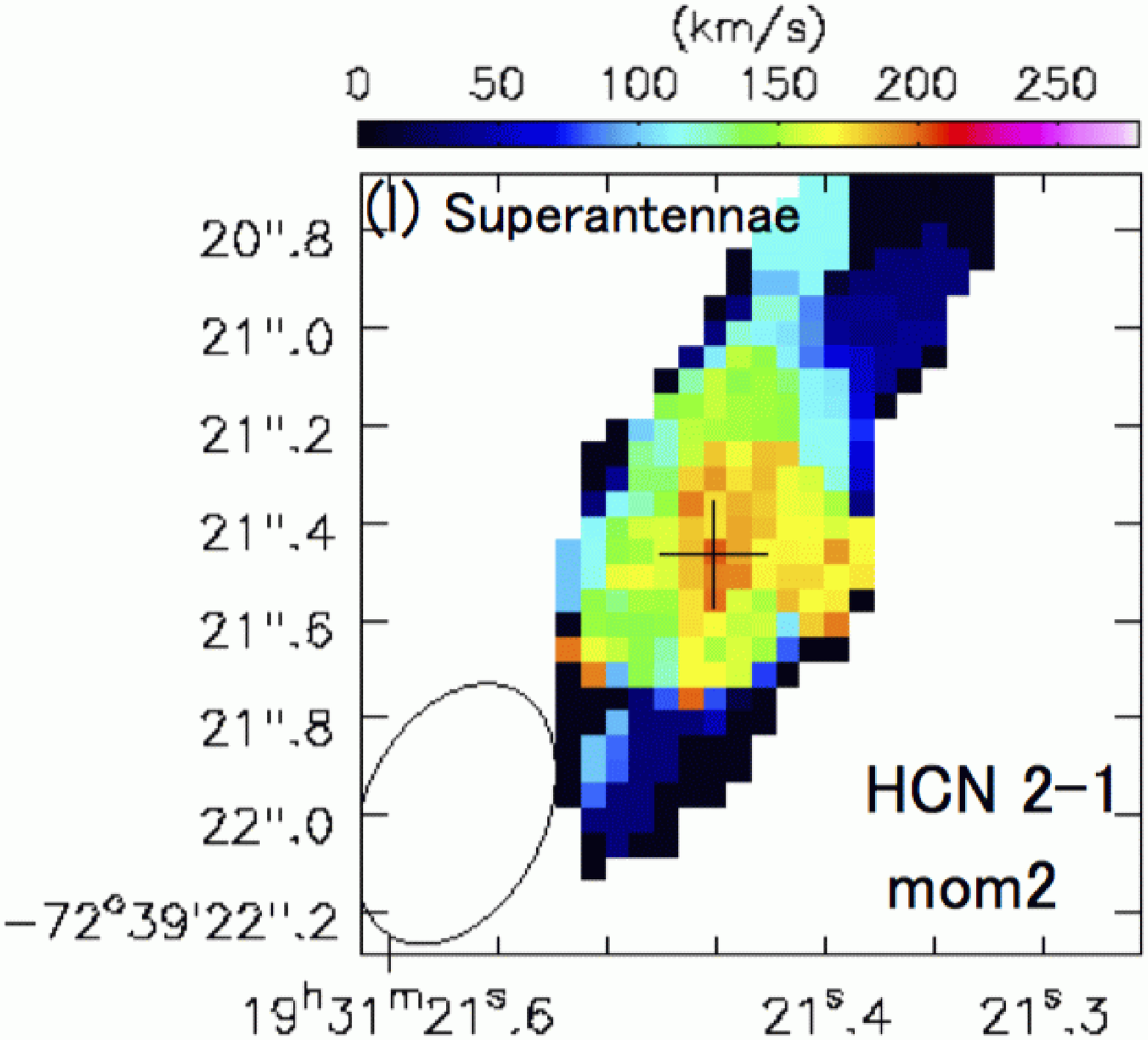} 
\end{center}
\caption{
{\it (Top)}: Integrated intensity (moment 0) map of 
(a) 183 GHz H$_{2}$O 
3$_{1,3}$--2$_{2,0}$, (b) HNC $J$ = 2--1, (c) HCO$^{+}$ $J$ = 2--1, and 
(d) HCN $J$ = 2--1 lines in ICRS coordinates.
Continuum emission simultaneously taken with individual lines is shown 
as contours (5, 15, and 25$\sigma$ with rms = 0.11 [mJy beam$^{-1}$] 
for the left two panels and 5 and 10$\sigma$ with 
rms = 0.21 [mJy beam$^{-1}$] for the right two).
The continuum peak flux is 3.4 (mJy beam$^{-1}$) (30$\sigma$) and 
3.1 (mJy beam$^{-1}$) (15$\sigma$) for the left and right pair of panels, 
respectively.
The vertical white bar in (a) corresponds to 1 kpc.
{\it (Middle)}: Intensity-weighted mean velocity (moment 1) map of 
the same lines.
{\it (Bottom)}: Intensity-weighted velocity dispersion (moment 2) map.
In (a)--(l), the continuum peak position is shown as a cross.
The filled or open circle in the lower left part indicates the synthesized 
beam size of each moment map.
An appropriate cutoff ($\sim$2$\sigma$) is applied to the moment 1 and 2 maps 
to prevent them from being dominated by noise.
}
\end{figure*}

\begin{table*}
\scriptsize
\caption{Molecular Emission Line Properties \label{tbl-1}}
\begin{center}
\begin{tabular}{lccl|cccc}
\hline
\hline
Line & \multicolumn{3}{c}{Integrated intensity (moment 0) map} & 
\multicolumn{4}{c}{Gaussian fit} \\  
 & Peak & rms & Beam & Velocity & Peak & FWHM & Flux \\ 
 & \multicolumn{2}{c}{[Jy beam$^{-1}$ km s$^{-1}$]} & 
[$''$ $\times$ $''$] ($^{\circ}$) & [km s$^{-1}$] & [mJy] & [km s$^{-1}$] &
[Jy km s$^{-1}$] \\  
(1) & (2) & (3) & (4) & (5) & (6) & (7) & (8) \\ \hline 
183 GHz H$_{2}$O & 4.3 (15$\sigma$) & 0.30 & 
0.50$\times$0.34 ($-$10$^{\circ}$) & 18205$\pm$14, 18738$\pm$20 $^{A}$
& 5.2$\pm$0.3, 4.9$\pm$0.2 $^{A}$ & 357$\pm$30, 562$\pm$46 $^{A}$ 
& 4.6$\pm$0.3 \\
HNC J=2--1 & 1.1 (5.0$\sigma$) & 0.23 & 
0.51$\times$0.35 ($-$9$^{\circ}$) & 18547$\pm$31 & 1.8$\pm$0.2 & 649$\pm$98 
& 1.2$\pm$0.2 \\
HCO$^{+}$ J=2--1 & 1.6 (6.3$\sigma$) & 0.29 & 
0.56$\times$0.37 ($-$26$^{\circ}$) & 18522$\pm$24 & 2.6$\pm$0.2 & 767$\pm$61 
& 2.0$\pm$0.2 \\
HCN J=2--1 & 2.9 (6.3$\sigma$) & 0.47 & 
0.57$\times$0.37 ($-$26$^{\circ}$) & 18564$\pm$17 & 3.8$\pm$0.2 & 810$\pm$48 
& 3.1$\pm$0.2 \\
\hline  
\end{tabular}
\end{center}

$^{A}$: Two Gaussian fit.

Notes:
Col. (1): Line.
Col. (2): Integrated intensity (in Jy beam$^{-1}$ km s$^{-1}$) at the 
emission peak. 
Detection significance relative to the rms noise (1$\sigma$) in the 
moment 0 map is shown in parentheses. 
Col. (3): Rms noise (1$\sigma$) level in the moment 0 map (in 
Jy beam$^{-1}$ km s$^{-1}$), derived from the standard deviation 
of sky signals in each moment 0 map. 
Col. (4): Beam size (in arcsec $\times$ arcsec) and position angle (in
degrees). The position angle is 0$^{\circ}$ along the north--south direction, 
and increases counterclockwise. 
Cols. (5)--(8): Gaussian fit of emission line in the spectrum at 
the continuum peak position, within the beam size. 
Col.(5): Optical local standard of rest (LSR) velocity (v$_{\rm opt}$) 
of the emission line peak (in km s$^{-1}$). 
Col. (6): Peak flux (in mJy). 
Col. (7): Observed full width at half maximum (FWHM) (in km s$^{-1}$).
Col. (8): Gaussian-fit, velocity-integrated flux (in Jy km s$^{-1}$). 
This flux will be used for our discussion of flux ratios.

\end{table*}

Figures 2a--2b show beam-sized spectra at the continuum peak, 
where the 183 GHz H$_{2}$O 3$_{1,3}$--2$_{2,0}$ line and HNC $J$ = 2--1, 
HCO$^{+}$ $J$ = 2--1, and HCN $J$ = 2--1 emission lines are clearly detected.
The 183 GHz H$_{2}$O emission line is characterized by a double-peaked 
profile, whereas the remaining three dense molecular tracers 
(HNC, HCO$^{+}$, and HCN $J$ = 2--1) display single-peaked emission line 
profiles.
We apply a double and single Gaussian fit for the 183 GHz H$_{2}$O and 
dense molecular lines, respectively.
Table 1 (columns 5--8) summarizes the fitting results.

\begin{figure*}
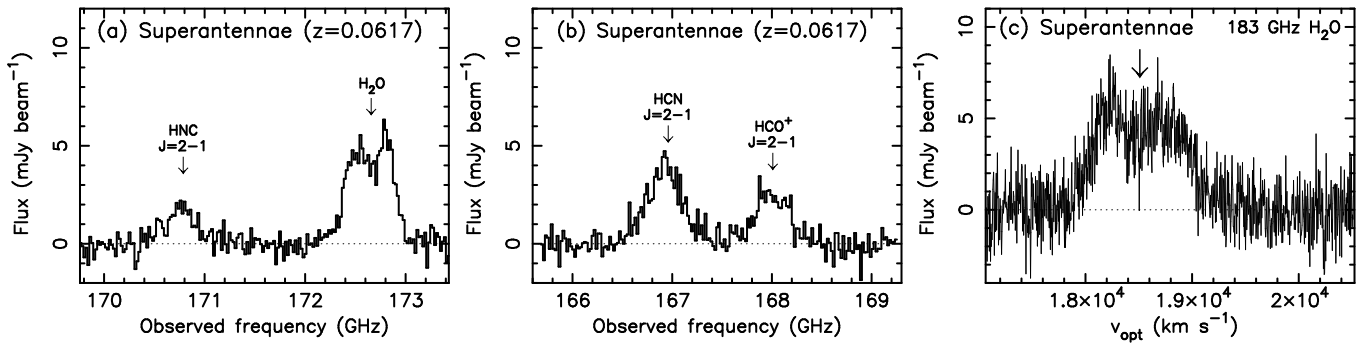

\begin{center}
\includegraphics[angle=-90,scale=.26]{f2a.eps} 
\includegraphics[angle=-90,scale=.26]{f2b.eps} 
\includegraphics[angle=-90,scale=.26]{f2c.eps} 
\end{center}
\vspace{-0.4cm}
\caption{
Spectrum of (a) 183 GHz H$_{2}$O 3$_{1,3}$--2$_{2,0}$ and 
HNC $J$ = 2--1 and 
(b) HCN $J$ = 2--1 and HCO$^{+}$ $J$ = 2--1 lines.
The abscissa is observed frequency (in GHz) and the ordinate is flux density 
(in mJy beam$^{-1}$).
The rms noise level is $\sim$1 mJy beam$^{-1}$ for $\sim$35 km s$^{-1}$ 
resolution.
(c) Detailed velocity profile of the 183 GHz H$_{2}$O 
3$_{1,3}$--2$_{2,0}$ emission line.
The abscissa is optical LSR velocity (in km s$^{-1}$) 
with $\sim$3.5 km s$^{-1}$ resolution and 
the ordinate is flux density (in mJy beam$^{-1}$).
The downward arrow indicates the systemic velocity (18510 km s$^{-1}$) 
of the Superantennae.
}
\end{figure*}

Intensity-weighted mean velocity (moment 1) and 
intensity-weighted velocity dispersion (moment 2) maps of 
183 GHz H$_{2}$O, HNC $J$ = 2--1, HCO$^{+}$ $J$ = 2--1, and 
HCN $J$ = 2--1 lines are displayed in Figures 1e--1l.
A similar rotation pattern is seen for the HNC $J$ = 2--1, 
HCO$^{+}$ $J$ = 2--1, 
and HCN $J$ = 2--1 lines, with the south-eastern side being blueshifted and 
the north-western side being redshifted (Figures 1f--1h), as previously seen 
for other dense molecular tracers, HCN $J$ = 3--2 and HCO$^{+}$ $J$ = 3--2 
\citep{ima16}, and near-infrared ($\sim$2 $\mu$m) emission lines 
\citep{reu07}.
However, the 183 GHz H$_{2}$O emission line shows no such rotation 
signature (Figure 1e), and yet shows much larger velocity dispersion 
values of $>$200 km s$^{-1}$ than the three dense molecular tracers 
at the nuclear ($\lesssim$1 kpc) region (Figures 1i--1l).
It is suggested that H$_{2}$O and dense molecular emission originate from 
dynamically and spatially different regions in the galaxy nucleus. 

\section{Discussion} 

In the Superantennae, the 183 GHz H$_{2}$O-to-HNC $J$ = 2--1 flux ratio is 
$\sim$4 (Table 1), which is by far the largest among $z <$ 0.15 ULIRGs, 
the ratios of which are $\lesssim$2 (M. Imanishi et al. 2021 in 
preparation).
The Superantennae is an outlier in this flux ratio.
HNC emission is sometimes observed to be weak in the vicinity of 
a luminous AGN if the column density of the surrounding molecular 
gas is insufficient for HNC shielding \citep[e.g.,][]{ima18,ima20}.
In fact, in the Superantennae, the flux of HNC, relative to HCN, 
at $J$ = 3--2 and $J$ = 4--3 is smaller than those of other ULIRGs 
\citep{ima18}.
To distinguish whether (1) the 183 GHz H$_{2}$O emission line is strong 
or (2) the HNC $J$ = 2--1 emission line is weak, we investigate 
the 183 GHz H$_{2}$O-to-HCO$^{+}$ $J$ = 2--1 flux ratio of the Superantennae 
and find it to be $>$2 (Table 1), which is also significantly higher than 
the ratios of $\lesssim$1 in the majority of other ULIRGs 
(M. Imanishi et al. 2021 in preparation).
The H$_{2}$O-to-HCN $J$ = 2--1 flux ratio of the Superantennae 
($>$1.4; Table 1) is also significantly higher than those of other ULIRGs 
($\lesssim$1; M. Imanishi et al. 2021 in preparation).
As HCN $J$ = 2--1 emission can be enhanced in AGNs 
\citep[e.g.,][]{koh05,ima07,kri08,ima09,izu16,ima16,ima19}, 
we thought that the excess H$_{2}$O-to-HCN $J$ = 2--1 flux ratio 
in the Superantennae may be weaker than other flux ratios, and yet 
it is still higher than those of other ULIRGs.
To summarize, in the Superantennae, the 183 GHz H$_{2}$O emission is 
significantly elevated relative to dense molecular line emission 
HNC $J$ = 2--1, HCO$^{+}$ $J$ = 2--1, and HCN $J$ = 2--1. 
Applying Equation (3) of \citet{sol05} to the observed emission line 
flux in Table 1 (column 8), the luminosity in units of 
(K km s$^{-1}$ pc$^{2}$) is estimated to be 
0.83 $\times$ 10$^{8}$, 1.4 $\times$ 10$^{8}$, and 2.3 $\times$ 10$^{9}$ 
for HNC $J$ = 2--1, HCO$^{+}$ $J$ = 2--1, and HCN $J$ = 2--1, respectively.
The probed nuclear dense molecular mass in the beam-sized spectrum 
(0$\farcs$55 $\times$ 0$\farcs$35 or 650 pc $\times$ 400 pc) is 
$>$a few $\times$ 10$^{8}$M$_{\odot}$, where we 
(1) use HCO$^{+}$ $J$ = 2--1 luminosity as the least biased dense 
molecular mass indicator (see the above-mentioned possible ambiguity 
for HNC and HCN around a luminous AGN), 
(2) assume that HCO$^{+}$ emission is optically thick and thermalized 
at $J$ = 2--1 and $J$ = 1--0 (i.e., same luminosity between $J$ = 2--1 and $J$ = 1--0 
in units of [K km s$^{-1}$ pc$^{2}$]), and 
(3) adopt a luminosity-to-mass conversion factor of 2--5 
(M$_{\odot}$ [K km s$^{-1}$ pc$^{2}$]$^{-1}$) for HCO$^{+}$ $J$ = 1--0 
\citep{ler17}.
The strong 183 GHz H$_{2}$O emission, even compared with the emission from 
such high-mass nuclear dense molecular gas, is remarkable.

We interpret that maser amplification of the 183 GHz H$_{2}$O emission 
line is a plausible physical mechanism for its flux elevation. 
First, while similar rotation patterns are seen in the moment 1 maps 
of the HNC $J$ = 2--1, HCO$^{+}$ $J$ = 2--1, and HCN $J$ = 2--1 lines 
(Figures 1f--1h), 
and so these emission are spatially resolved, the lack of such 
rotation patterns (Figure 1e) suggests that the 183 GHz H$_{2}$O emission 
originates in very compact, spatially unresolved molecular gas.
We apply the CASA task ``imfit'' to moment 0 maps (Figures 1a--d) 
and find that the deconvolved intrinsic emission size of 
H$_{2}$O ($\lesssim$180 mas or $\lesssim$220 pc) is much smaller than 
those of HCN, HCO$^{+}$, and HNC $J$ = 2--1 (450--1050 mas or 550--1250 pc).
This is also confirmed by visibility fitting, using the CASA 
task ``uvmodelfit''.  
AGN-illuminated dense and warm molecular gas with coherent velocity 
at the very centre ($<<$1 kpc) of the nucleus is the most plausible site for 
maser amplification \citep{neu94,mal02} and can naturally explain the 
compact, spatially unresolved 183 GHz H$_{2}$O emission.

Second, unlike dense molecular tracers with single-peaked emission 
line profiles, the 183 GHz H$_{2}$O emission line is double-peaked.
This double-peaked emission line profile, by definition, can produce 
large velocity dispersion values in the moment 2 map of the H$_{2}$O line 
(Figure 1i). 
Figure 2c shows the detailed profile of the H$_{2}$O emission line 
with finer velocity resolution ($\sim$3.5 km s$^{-1}$), where 
the blueshifted and redshifted emission components are brighter 
than the systemic velocity component, as often seen in the bright 
22 GHz H$_{2}$O megamaser emission in AGNs 
\citep[e.g.,][]{bra96,gre03b,bra04,hen05,kon06a,kon06b}. 
The observed double-peaked emission line profile of the Superantennae 
can naturally be reproduced if maser-origin blueshifted and redshifted 
H$_{2}$O emission in the vicinity of the mass-accreting SMBH are superposed 
onto very broad (FWHM $\sim$ 650--800 km s$^{-1}$; Table 1) thermal H$_{2}$O 
emission from the galaxy nucleus ($\sim$1 kpc).
A compact rotating maser disc at both sides of the central 
mass-accreting SMBH can produce stronger blueshifted and 
redshifted emission than the systemic velocity component 
at the foreground side of the SMBH \citep{miy95}.
The 183 GHz H$_{2}$O emission detected in other $z <$ 0.15 ULIRGs 
shows a single-peaked line profile, similar to those of dense 
molecular tracers (M. Imanishi et al. 2021 in preparation).
To summarize, the luminous, spatially compact, spectrally double-peaked 
183 GHz H$_{2}$O emission in the Superantennae can naturally be 
explained by maser amplification in AGN-illuminated gas at the very 
centre of the galaxy nucleus, and an enhanced H$_{2}$O abundance in 
a large volume of nuclear ($\sim$1 kpc) dense molecular gas is 
disfavoured as the origin of the elevated 183 GHz H$_{2}$O emission. 
While a large fraction of AGNs with detected 22 GHz H$_{2}$O megamaser 
emission show multiple narrow features, the maser-origin 183 GHz 
H$_{2}$O emission of the Superantennae is broad, as seen in a few AGNs for 
the 22 GHz line \citep[e.g.,][]{koe95,bra96}.
We interpret that H$_{2}$O maser spots in the Superantennae distribute 
widely and continuously along the radial direction from the central 
mass-accreting SMBH. 

The observed 183 GHz H$_{2}$O emission line properties in the 
Superantennae 
are significantly different from those of another ULIRG with 183 GHz 
H$_{2}$O emission line detection, Arp 220 ($z =$ 0.018 or $\sim$80 Mpc) 
\citep{cer06,gal16,kon17}.
In Arp 220, the 183 GHz H$_{2}$O emission line flux is smaller than 
those of the HNC, HCO$^{+}$, and HCN $J$ = 2--1 lines \citep{gal16}, similar to 
the majority of $z <$ 0.15 ULIRGs 
(M. Imanishi et al. 2021 in preparation).
The line profile of the 183 GHz H$_{2}$O emission in Arp 220 is also 
single-peaked in a similar way to those of dense molecular lines 
\citep{cer06,gal16,kon17}.
These observational results of the 183 GHz H$_{2}$O emission in Arp 220 
can naturally be explained by thermal emission.  
The Superantennae shows the strongest AGN-origin megamaser emission 
signatures for the 183 GHz H$_{2}$O line beyond the immediately local 
universe at $>$20 Mpc.

Adopting Equation (1) of \citet{sol05}, we obtain the 183 GHz H$_{2}$O 
maser (isotropic) luminosity 
$L_{\rm 183GHz-H_{2}O}$ $\sim$ 6.1 $\times$ 10$^{4} L_{\odot}$.
Even though a fraction of this H$_{2}$O luminosity comes from thermal 
emission in the nuclear region ($\sim$1 kpc) of the galaxy, the fact 
that the H$_{2}$O-to-HCO$^{+}$ $J$ = 2--1 and H$_{2}$O-to-HNC 
$J$ = 2--1 flux ratios in the Superantennae are $\gtrsim$2 times higher 
than those in other $z<$ 0.15 ULIRGs suggests 
that at least half the luminosity ($>$a few $\times$ 10$^{4}L_{\odot}$) 
originates from maser emission from the AGN-illuminated gas ($<<$1 kpc).
This is as high as or possibly even higher than the most luminous 
22 GHz H$_{2}$O megamaser emission in an AGN at $z \sim$ 0.66 
($\sim$2 $\times$ 10$^{4}L_{\odot}$) \citep{bar05}.
The Superantennae is a promising target for detecting multiple H$_{2}$O 
megamaser emission lines to constrain the physical properties of 
AGN-illuminated 
dense and warm molecular gas around an actively mass-accreting SMBH. 
The upper and lower energy levels of H$_{2}$O lines are 
($E_{\rm u}$, $E_{\rm l}$) = (643 K, 642 K) for the 6$_{1,6}$--5$_{2,3}$ 
transition at 22 GHz, and (205 K, 196 K) for the 3$_{1,3}$--2$_{2,0}$ transition 
at 183 GHz.
Depending on molecular gas temperature, the 183 GHz to 22 GHz H$_{2}$O 
maser flux ratio can vary widely \citep{cer06}.
For the Superantennae, no detection was reported for the well-studied 
22 GHz H$_{2}$O maser emission with shallow (rms $\gtrsim$ 75 mJy) 
observations \citep{gre02}.
More sensitive 22 GHz observations 
will provide important insights into the properties of AGN-illuminated gas 
at the very centre of galaxy nuclei in the distant universe 
at $>$270 Mpc.

\section{Summary} 

We conducted ALMA band-5 observations of the Superantennae galaxy 
at $z =$ 0.0617. 
We detected remarkably strong 183 GHz H$_{2}$O 3$_{1,3}$--2$_{2,0}$ 
emission, relative to dense molecular lines 
(HNC, HCO$^{+}$, and HCN $J$ = 2--1) when compared to other ULIRGs. 
The observed compact, spatially unresolved emission nature 
and double-peaked emission line profile of H$_{2}$O   
are markedly different from the spatially resolved, single-peaked 
emission line properties of dense molecular tracers in the 
galaxy nucleus.
We argue that the detected luminous 183 GHz H$_{2}$O emission 
originates in maser amplification in AGN-illuminated molecular gas 
around a mass-accreting SMBH at the very centre of the galaxy nucleus.
This pushes (sub)millimetre H$_{2}$O maser emission detection 
up to $>$270 Mpc, far beyond the immediately local universe, 
owing to ALMA's high sensitivity.
Combination with so-far undetected 22 GHz H$_{2}$O maser emission will 
constrain the physical properties of dense and warm molecular gas 
in the vicinity of a mass-accreting SMBH.

\vspace*{-0.7cm}

\section*{Acknowledgements}
We thank the referee for valuable comments.
This paper makes use of the following ALMA data: 
ADS/JAO.ALMA\#2017.1.00022.S.
Data analysis was in part carried out on the open use data analysis
computer system at the Astronomy Data Center, ADC, of the National
Astronomical Observatory of Japan. 

\vspace*{-0.7cm}

\section*{Data Availability}

The data used in this paper (2017.1.00022.S) 
are available in the ALMA archive at
https://almascience.eso.org/.

\vspace*{-0.6cm}



\bibliographystyle{mnras}
{}




\bsp	
\label{lastpage}
\end{document}